\documentclass{article}

\usepackage{PRIMEarxiv}

\usepackage[utf8]{inputenc} 
\usepackage[T1]{fontenc}    
\usepackage{hyperref}       
\usepackage{url}            
\usepackage{booktabs}       
\usepackage{amsfonts}       
\usepackage{nicefrac}       
\usepackage{microtype}      
\usepackage{lipsum}
\usepackage{fancyhdr}       
\usepackage{graphicx}       
\graphicspath{{media/}}     
\usepackage{caption}
\usepackage{subcaption}
\usepackage{ptdr-definitions}  
\usepackage{heppennames2}      
\usepackage[
backend=biber,
sorting=none
]{biblatex}
\renewbibmacro{in:}{}
\usepackage{amsmath,amsfonts,bm}
\usepackage{ amssymb }

\def\eqref#1{equation~\ref{#1}}

\def\1{\bm{1}}

\DeclareMathAlphabet{\mathsfit}{\encodingdefault}{\sfdefault}{m}{sl}
\SetMathAlphabet{\mathsfit}{bold}{\encodingdefault}{\sfdefault}{bx}{n}

\usepackage{mathtools}

\title{Electron Energy Regression in the CMS High-Granularity Calorimeter Prototype}
\author{}

\begin{document}
\maketitle
\vspace{-25mm}

\begin{center}
    Roger Rusack$^{1\dagger}$,
    Bhargav Joshi$^{1\dagger}$,
    Alpana Alpana$^{2\dagger}$,
    Seema Sharma$^{2\dagger}$,
    Thomas Vadnais$^{1\dagger}$
    \\
    \vspace{3mm}
    \textsuperscript{1} Department of Physics, University of Minnesota, Minneapolis, USA
    \\
    \textsuperscript{2} Department of Physics, IISER-Pune, India
    \\
    \vspace{3mm}
    \textsuperscript{\dag} These authors contributed equally to this work 
    \\
\end{center}
\vspace{10mm}

\begin{abstract}

We present a new publicly available dataset that contains simulated data of a novel calorimeter to be installed at the CERN Large Hadron Collider. This detector will have more than six-million channels with each channel capable of position, ionisation and precision time measurement. Reconstructing these events in an efficient way poses an immense challenge which is being addressed with the latest machine learning techniques. As part of this development a large prototype with 12,000 channels was built and a beam of high-energy electrons incident on it. Using machine learning methods we have reconstructed the energy of incident electrons from the energies of three-dimensional hits, which is known to some precision. By releasing this data publicly we hope to encourage experts in the application of machine learning to develop efficient and accurate image reconstruction of these electrons.

\keywords{HGCAL \and FAIR Data \and Energy Regression \and Machine Learning \and DNN}

\end{abstract}

\section{Introduction}

\begin{figure}[htbp]
    \centering
    \includegraphics[width=0.85\textwidth]{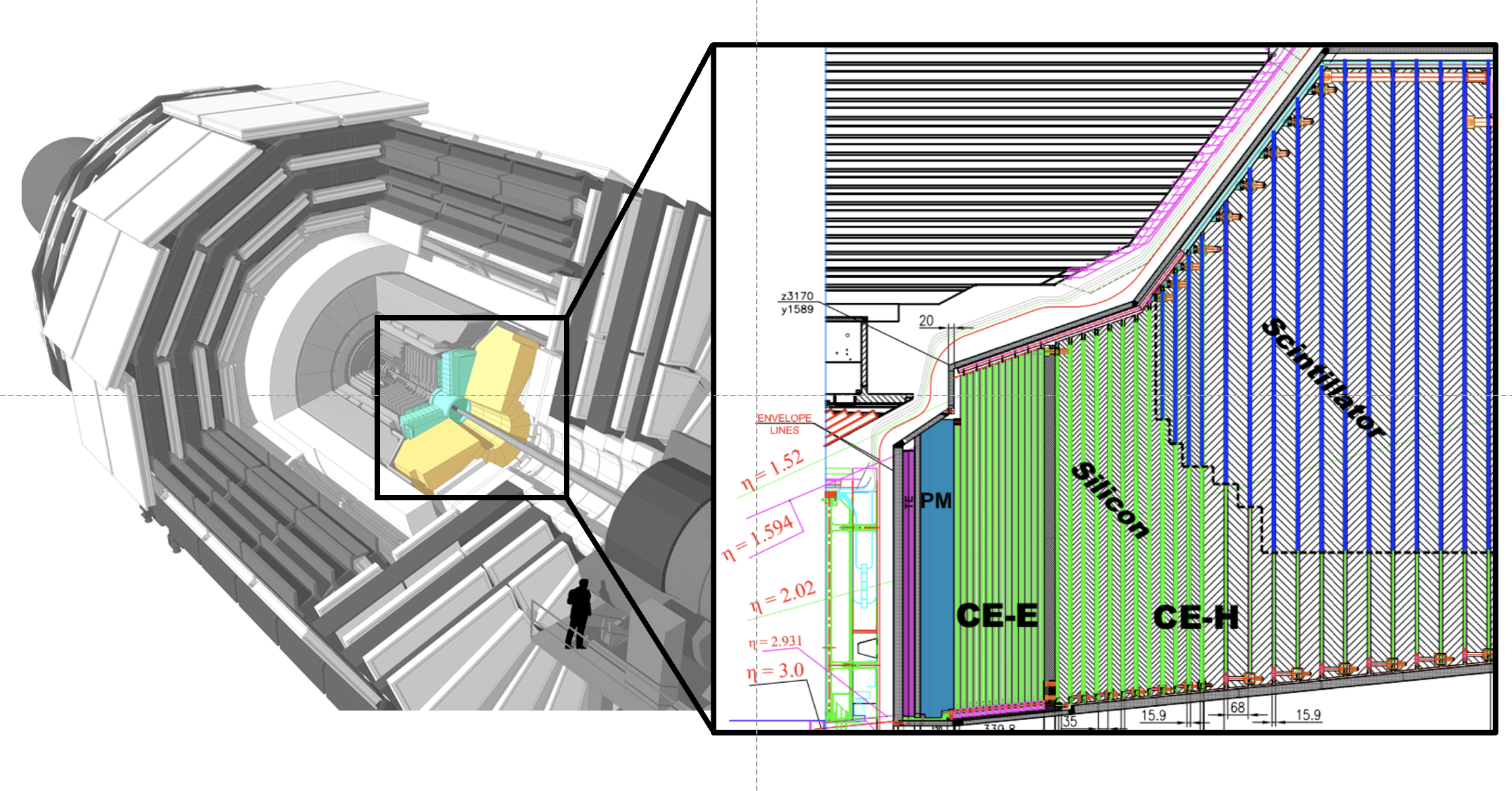}
    \caption{Current design of the CMS detector (left) to the human scale. The highlighted regions in blue and yellow color represent the ECAL and the HCAL detectors. These regions will be replaced by the newly designed calorimeter (right). It consists of three successive layers which combine the functionalities of both, the ECAL and the HCAL.}
    \label{fig:hgcal_upgrade}
\end{figure}

To measure the energy of particles produced in collisions at the large hadron collider (LHC) the Compact Muon Solenoid (CMS) experimental detector currently has in each of its two endcaps an electromagnetic calorimeter (ECAL), equipped with a preshower (ES) detector, and a hadronic calorimeter (HCAL). Between the interaction point (IP) where the collisions occur there is a silicon tracking detector to measure the momentum of charged particles as they move through the solenoidal magnetic field. Towards the end of this decade the LHC will be upgraded to the High-Luminosity LHC (HL-LHC) where the collision rate of the colliding beams will be increased by a factor of three or more. To cope with the high radiation levels from the particles produced in the collisions the calorimeters in the endcaps will be replaced with a new type of calorimeter, the high-granularity calorimeter (HGCAL), which tracks the progression of the loss of energy by high energy particles by sampling of the shower at different depths inside it. The HGCAL will be constructed from radiation hard silicon sensors, or plastic scintillator sensors, where the radiation levels are lower, that are sandwiched between passive layers of absorber material made of steel or lead. The location within the CMS detector and an outline of the design are shown in Fig. \ref{fig:hgcal_upgrade}.\\
\newline
In the HGCAL there will be approximately three million detector channels in each of the two endcaps. The information of the energy deposited by particles and the time of their arrival in each channel is measured and digitized.  This information is transmitted to off-detector electronics for processing and storage. 
How this information is used to reconstruct the energy of as incident electron, it's impact on the calorimeter and its angle of incidence is a challenge that we discuss in this paper.  In calorimetry the typical method to reconstruct of electrons is with seeding and clustering methods. With the HGCAL design\ref{fig:hgcal_upgrade}, which has considerably more information available than in earlier examples of calorimeters, new algorithms based on modern machine learning (ML) methods can be developed to solve the reconstruction problem, which in a sense is like a three-dimensional image reconstruction problem.  In this paper we discuss the problem of reconstructing high-energy electrons from the energy deposits in the sensors in the HGCAL. 

For this we have generated a large volume of simulated data using the GEANT4\cite{geant4} simulation package, which accurately simulates electromagnetic showers generated by electrons impacting the calorimeter. This data is available at Zenodo\footnote{\href{https://zenodo.org/}{https://zenodo.org/}} and can be used to test new ML methods to address this problem. To accompany the data we provide exemplar software and metadata to permit non-specialist access to the data and development of novel solutions. The exemplar software describes how to access the data and provides a simple reconstruction example that is based on a Deep Neural Network (DNN). In this paper we describe the problem to be solved in more detail and the results that we have obtained with the DNN model.

 \section{The High Granularity Calorimeter}

The entire assembly of each of the two HGCAL calorimeters weighs approximately 230 T and will be used to measure the energies of particles produced at the IP with angles of approximately 10 to 30 degrees from the beam axis
\footnote{The coverage is between 1.5 and 3.0 in pseudorapidity defined as $\eta=-ln|\tan\dfrac{\theta}{2}|$, where $\theta$ is the azimuthal angle relative to the beam axis.}. In the final detector the first 26 layers will form the electromagnetic (CE-E)~\cite{cee_commissioning} section which will have hexagonal silicon sensors of about 8" width divided into hexagonal cells with areas of 1.1 and 0.5 cm$^2$. Behind the CE is the 21-layer hadronic section (CH). In this the first eight layers will consist of silicon sensors similar to the CE-E section, and the last 12 layers will have a mixture of silicon sensors and plastic scintillators. 


\subsection{The Prototype Setup}

\begin{figure}[htbp]
	\centering
	\includegraphics[width=5in]{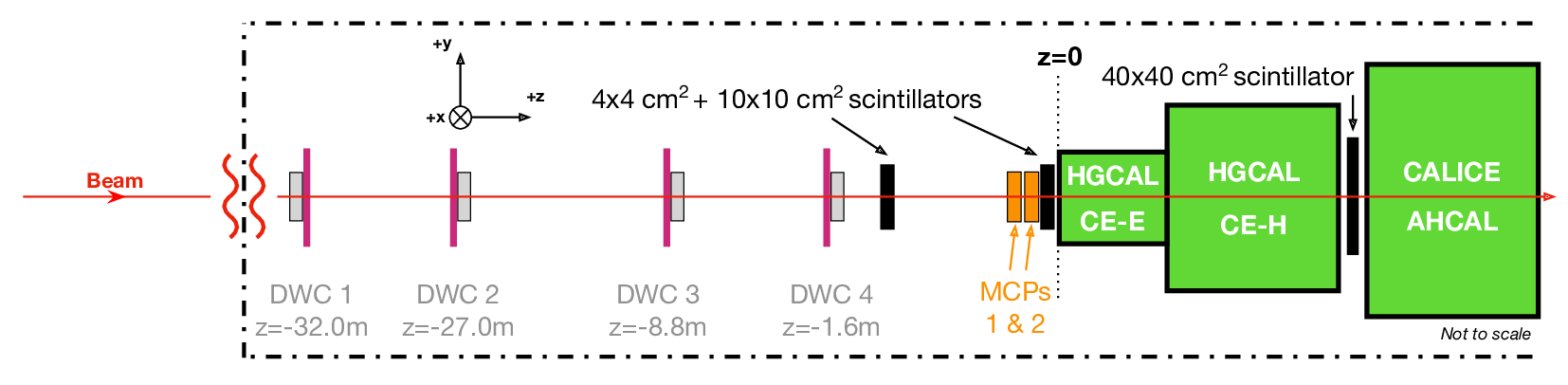}
	\caption{\label{fig:hgcal_tb_setup} The test beam setup of the prototype along the H2 beam line. The four delay wire chambers (DWCs) track the position of the incoming positron. For triggering on signal events two plastic scintillators and fast multiplying tubes are used.}
\end{figure}

\begin{figure}
    \centering
    \includegraphics[width=0.35\textwidth]{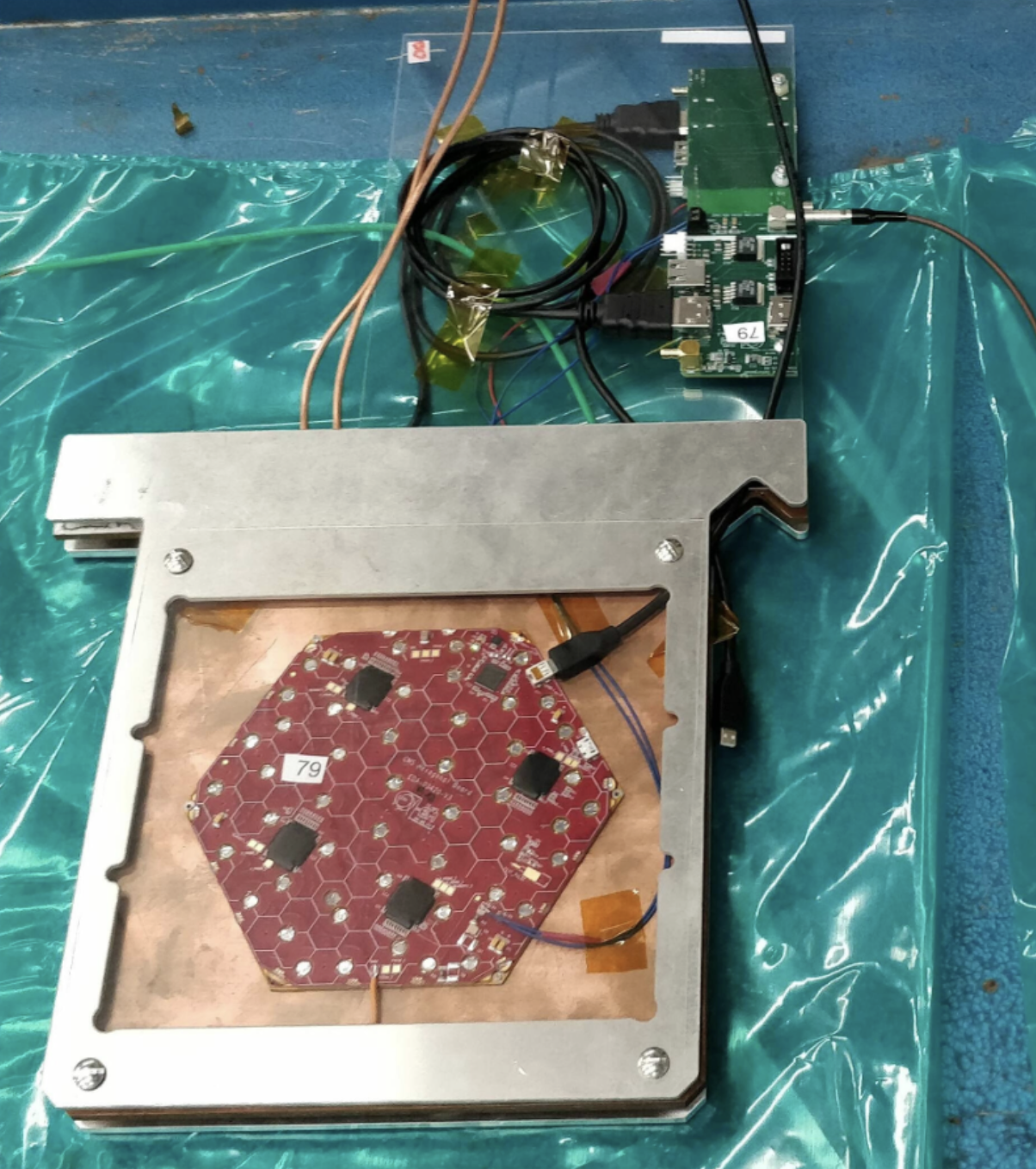}
    \includegraphics[width=0.64\textwidth]{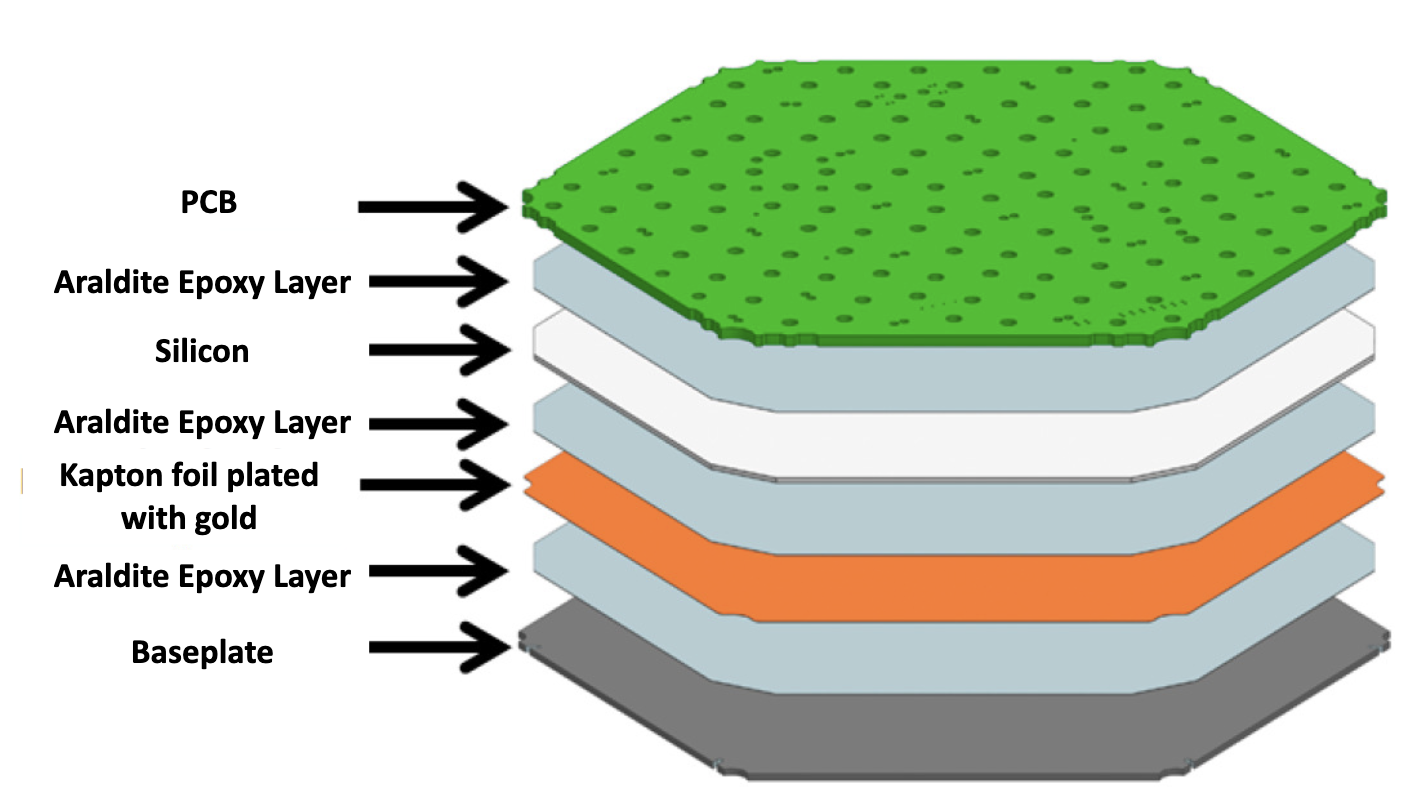}
    \caption{A front view of a prototype of the CE-E minicassette (left). It consists of two Hexagonal module mounted onto a Cu cooling plate on either side.  The module is an assembly (right) of a baseplate
    made of copper or copper-tungsten, a 100 $\mu$m thick gold-plated Kapton$^\circledR$ sheet, a hexagonal silicon sensor, and a printed circuit board called 'hexaboard'. Araldite$^\circledR$ is used an an expoxy to glue different components in the module.}
    \label{fig:Minicassette_image}
\end{figure}

To evaluate the performance of the detector and to qualify many aspects of the design a large-scale prototype of the HGCAL was built and tested in the H2 beamline at CERN's Pr\'evessin site (Figure~\ref{fig:hgcal_tb_setup}). A beam of positrons is provided by Super Proton Synchrotron (SPS) accelerator. Since, the positron is an anti-particle of the electron differing only in electric charge, the response of the interaction of positron in the prototype is same as that of an electron without any external magnetic field. The prototype consisted of 3 sections, Electromagnetic (CE-E), Hadronic (CE-H) and a CALICE Analog Hadronic Calorimeter (AHCAL)\cite{calicecollaboration2022design,calice_1}, arranged in series in that order. This is similar to the final configuration of the HGCAL. The CE-E~\cite{hgcal_tb_e_nov22} section consists of 28 sampling layers made using 14 double-sided mini-cassettes (Figure~\ref{fig:Minicassette_image} right). Each cassette consist of a lead clad with stainless steel or Cu/CuW absorber sandwiched between two silicon sensor layers. The hexagonal silicon sensors are subdivided into 128 hexagonal silicon detector channels. Each channel is equipped with electronics to measure the energy and the time of the particle interactions in the sensor. The entire CE-E section corresponds to a total of 26 radiation lengths or 1.4 nuclear interaction lengths.

In the prototype the CE-H~\cite{hgcal_tb_pi_nov22} section was composed of 12 sampling layers each with seven Si modules arranged in daisy structure, each layer was sandwiched between a 40~mm thick steel plate. Due to the limited availability of silicon sensor modules, the last three layers of CE-H were equipped with only one sensor module placed at the center of the layer.The CE-H is followed by a 4.4 nuclear interaction length deep prototype of the AHCAL that was built with 39 sampling layers of SiPM-on-scintillator-tile active layers interspersed between steel absorbers.
\section{Electromagnetic Showers}
 
\begin{figure}[htbp]
	\centering
	\includegraphics[width=4in]{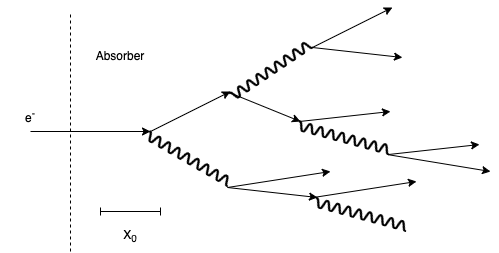}
	\caption{A schematic showing the development of an electromagnetic shower by an incoming electron in an absorber.}
	\label{fig:em_shower}
\end{figure}

 When energetic particles pass through a media, they typically loses energy through coulomb interactions with the electrons in the media. Energetic electrons (E ~ 1 GeV), on the other lose energy primarily via emission of \textit{bremsstrahlung} radiations. When the electron passes through a dense media, it get accelerated or decelerated quickly due to the strong electric fields of the nuclei which causes it to emit radiations or photons. Energetic photons, on the other hand, produce pairs of electrons and positrons as they interact with the nuclei of the atoms. This results in a cascade of secondary particles know as an Electromagnetic Shower\ref{fig:em_shower} and the process continues until the energy of the decay products falls below a critical energy E$_c$.

 These showers can be characterized by several parameters, which include the \textit{radiation length} and $\textit{Moli\`ere radius}$. The \textit{radiation length} is defined as the distance over which the energetic electron loses 1/e fraction of its energy. Thus, the "shower depth" can be written in terms of the \textit{radiation length} as follows

 \begin{equation}
 X = X_0\dfrac{\ln(E/E_c)}{\ln(2)}
 \label{eqn:energy_res}
 \end{equation},

 where E$_c$ is the critical energy\footnote{\href{https://pdg.lbl.gov/2022/AtomicNuclearProperties/critical_energy.html}{https://pdg.lbl.gov/2022/AtomicNuclearProperties/critical\_energy.html}} of electron in a given material.

As the electron dissipates energy, the size of the spread increases in directions orthogonal to its momentum. The the $\textit{Moli\`ere}$ radius can be used to define the lateral spread of the shower till the critical energy reached as the electron traverses $X_0$ through the medium. By definition, a cylinder of $\textit{Moli\`ere radius}$ contains about 90\% of the total deposited energy.

 The electromagnetic calorimeters are designed to capture the highly energetic photons and electrons and measure their energies. They can also localise the position of the incoming particle in space and, in some cases, measure its direction. The part of the calorimeter that produces showers is known as the absorber material, whereas, the material that measures the energy is known as the active part. Ideally a calorimeter has a small $X_0$ and $\textit{Moli\`ere}$ radius to contain the showers as effectively as possible. The electromagnetic calorimeters can either be of homogeneous type or of sampling type. Homogeneous calorimeters typically have one block of absorber, where the incoming particle dissipates energy and the active material surrounding it measures the energy. In a sampling calorimeter, there are alternating layers of absorbers and active materials, and the energy dissipated in one layer is measured using the energy deposited in the layers before and after the absorber. Finally, the sum of energies over all the layers gives the total energy deposited can be used to measure the energy of the incoming particle.

The energy resolution of a calorimter gives its precision in measuring the energy. For an electromagnetic calorimeter, the energy resolution can be written as follows.

\begin{equation}
\dfrac{\sigma}{E}=\dfrac{S}{\sqrt{E}}\oplus\dfrac{N}{E}\oplus C,
\label{eqn.energy_res}
\end{equation}

where the first term on the right-hand side is the \textit{stochastic} or \textit{sampling} term, the middle term is the \textit{noise} term and the last term is the \textit{constant} term. The \textit{stochastic} term arises from the fact that the number of primary and secondary particles produced in the interactions fluctuates. The \textit{noise} term, on the other hand, comes from the noise in the detector electronics. Furthermore, this term receives contributions from other simultaneous interactions or collisions happening in the same event known as "pileup". Finally, the constant term is the measure of quality of the detector construction. It accounts for the imperfections in the geometry, non-uniformity in the response and energy losses that cannot be measured by its electronics.
\section{Dataset}

The dataset consists of simulations of reconstructed hits, known as "rechits", produced by positrons passing through the HGCAL test beam prototype. For simulations, Monte Carlo method is used to produce  positrons with energy ranging from 10 to 350 GeV. In the next step, GEANT4~\cite{geant4} package is used to simulate their interactions with the detector material. The conditions used in generating positrons are fine tuned to account for real detector effects such as energy losses in the beam. The simulated hits are then digitized using the CMS software. The digitized information was then processed through the CMS software to reconstruct the signals as hits within the detector. The rechits along with their details pertaining to signal reconstruction was stored in root~\cite{cern_root} format. These files were then skimmed using uproot~\cite{jim_pivarski_2021_4537826} package to obtain the final dataset. A set of preselections is applied to ensure that the event selection is identical to the one used in performing the analysis~\cite{hgcal_tb_e_nov22} published by the CMS collaboration. The hits are chosen to have a minimum energy of 0.5 MIP\footnote{Minimum Ionizing Particle (MIP) is the unit used to count the energy of digitized hits.}, which is well above the HGCAL noise levels. Events with more than 50 hits in CE-H layers are rejected. The track of electron extrapolated using the hits from the DWC chambers is required to be within a 2x2 cm$^2$ window within the first layer. The final dataset is a set of 3.2 million events, each event containing position coordinates of rechits within the detector and their calibrated energies. \href{https://portal.hdfgroup.org/display/HDF5/HDF5}{HDF5} format is used to organize the data in hierarchical arrays. The file contains following the arrays:
\begin{itemize}
    \item \textbf{nhits}: An integer array representing number of reconstructed hits (rechits) in each event.
    \item \textbf{rechit\_x}: A nested array of length equal to the number of events and sub-arrays of length of nhits. Each sub-array contains a floating value representing x-coordinate of the position of the rechits in units of centimeters.
    \item \textbf{rechit\_y}: A nested array with a structure and size same as rechit\_x. Each floating value represents the y-coordinate of the position of a rechit in units of centimeters.
    \item \textbf{rechit\_z}: A nested array with a structure and size same as rechit\_x. Each floating value represents the z-coordinate of the position of a rechit in units of centimeters.
    \item \textbf{rechit\_energy}: A nested array with a structure and size same as rechit\_x. Each floating value represents the calibrated energy of a rechit in units of MIPs.
    \item \textbf{target}: The true energy of the incoming positron in units of GeV.
\end{itemize}

To ensure the FAIR-ness of the publication of the dataset, it has been published~\cite{bhargav_joshi_2023_7504164} on Zenodo~\cite{zenodo} platform, which was launched in May 2013 as part of the \href{https://explore.openaire.eu}{OpenAIRE project}, in partnership with CERN. 
The dataset\cite{bhargav_joshi_2023_7504164} consists of two files in \textit{gzip} format. These can be uncompressed to obtain two files in HDF5 format. The smaller sample of 648,000 events with a label "0001" has a file size of 2.8 GB and the full dataset with a label "large" has a file size of 14.0 GB. The code to unpack and use the dataset has been made available on Github\footnote{\href{https://github.com/FAIR-UMN/FAIR-UMN-HGCAL}{https://github.com/FAIR-UMN/FAIR-UMN-HGCAL}}. The metadata describing the contents of the file are available in JSON format under the same repository.
\section{Summary}

\begin{figure}[!ht]
\centering
\includegraphics[width=0.75\textwidth]{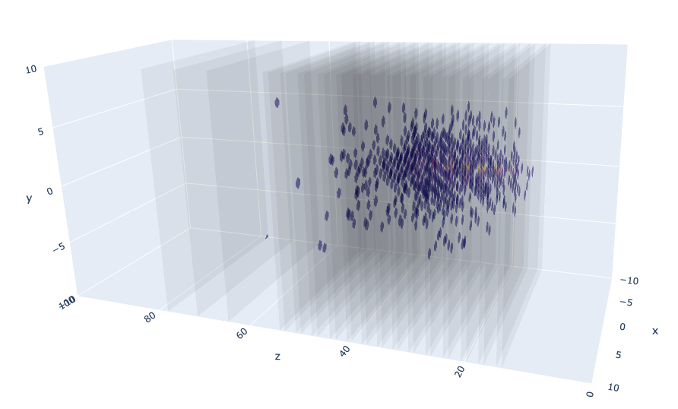}
\caption{\label{fig:event_display} An event display of a simulated event of a 100 GeV positron passing through the prototype. The energy of reconstructed hits is measured in the units of Minimum Ionizing Particles (MIPs).}
\end{figure}

The purpose of the release of the dataset is to make it open for everyone for building models for estimating the resolution with better precision, develop visualization tools and benchmarking ML techniques such as Generative Adversarial Networks (GANs), which can be used for generating EM showers with reduced computational time. For the purpose of exploring the dataset, the source code of the simple DNN model that was developed in python for energy regression has been added to the aforementioned Github repository. The respository has been built using the using the "cookiecutter" template used by the FAIR4HEP group for ensuring Findability and reproducibility of the results. An example notebook in the repository also demonstrates a way to make event displays (Figure~\ref{fig:event_display}) of individual events in the dataset.

\begin{figure}[!ht]
\centering
     \begin{subfigure}[b]{0.45\textwidth}
        \centering
        \includegraphics[width=1.0\textwidth]{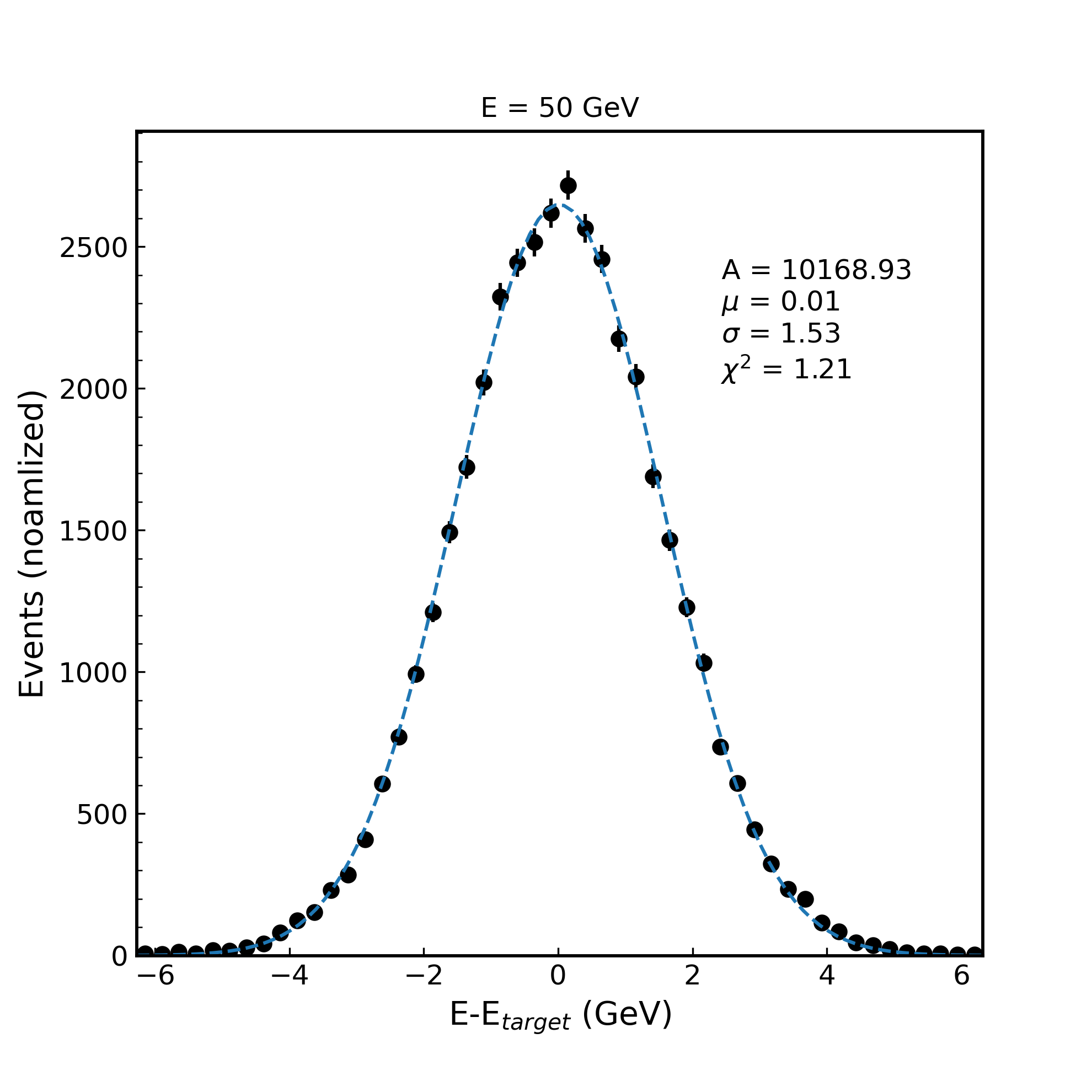}
     \end{subfigure}
     \begin{subfigure}[b]{0.45\textwidth}
        \includegraphics[width=1.0\textwidth]{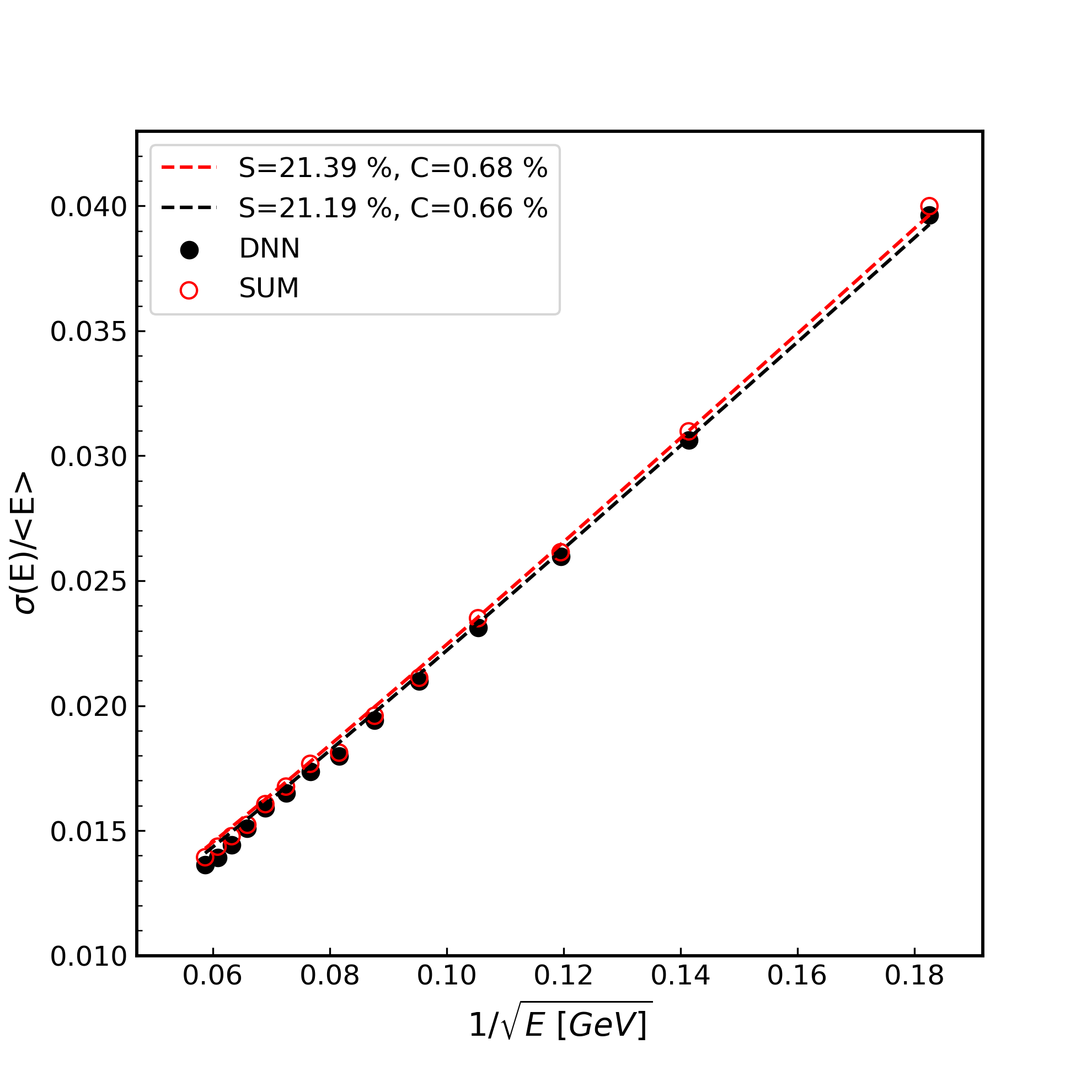}
     \end{subfigure}
    \caption{Predicted energies of positrons between a range of [40, 60] GeV particle as predicted by the DNN relative to the energy of the incoming particle (left). The data points are fitted with a Gaussian distribution using a minimum $\chi^2$ fit. Resolution plotted as a function of the inverse of square root of the energy of the simulated particle (right). The resolutions obtained through summing the energies of rechits (black line) and those obtained through the output of DNN (red line) are comparable.}
    \label{fig:resolution_vs_sqrtE}
\end{figure}

After training on the simulated dataset using a fully connected DNN, the performance of the network can be evaluated by computing the energy resolution in different bins of energies. To achieve this, the difference between measured and true energies from the simulations are plotted for energies ranging from 20 to 300 GeV in 14 bins of 25 GeV width. In each bin, the resulting distribution has a shape of a Gaussian distribution. This distribution is then fit using a $\chi^2$ minimization technique to obtain the mean and the variance. The mean represents the bias in the estimation in each bin, whereas the ratio of the variance to the mean gives the estimate of the energy resolution. Without any contributions from pileup, the \textit{noise} term in (Equation~\ref{eqn:energy_res}) is assumed to be zero. The squares of the resolutions obtained from the 14 energy bins can be fitted as the sum of quadratures of the \textit{stochastic} term and the constant term. The slope and the intercept of the linear fit (Figure~\ref{fig:resolution_vs_sqrtE}) provides an estimate for the \textit{stochastic} term and the constant term respectively.

\section{Acknowledgements}
This work has been supported by the Department of Energy, Office of Science, Office of Advanced Scientific Computing under award number DE-SC0021395. 
The authors would like to express their gratitude to the CMS Collaboration, and in particular to the CMS HGCAL community for making the providing the configurations files to generate simulated events. We would also like to thank our colleagues from the FAIR4HEP group for discussions and their invaluable inputs and suggestions for writing this paper.
\newpage
\printbibliography
\end{document}